\documentclass[conference]{IEEEtran}
\IEEEoverridecommandlockouts

\usepackage{cite}
\usepackage{amsmath,amssymb,amsfonts}
\usepackage{algorithmic}
\usepackage{graphicx}
\usepackage{textcomp}
\usepackage{xcolor}
\usepackage{booktabs}
\usepackage{url}
\usepackage{hyperref}
\usepackage{listings}
\usepackage{multirow}
\usepackage{tikz}
\usepackage[section]{placeins}

\hypersetup{
  colorlinks=true,
  linkcolor=blue,
  citecolor=blue,
  urlcolor=blue
}

\newcommand{\safeincludegraphics}[2][]{%
\IfFileExists{#2}{%
\includegraphics[#1]{#2}%
}{%
\fbox{\parbox{0.92\linewidth}{\centering Missing figure:\\ \texttt{\detokenize{#2}}}}%
}%
}

\def\BibTeX{{\rm B\kern-.05em{\sc i\kern-.025em b}\kern-.08em
    T\kern-.1667em\lower.7ex\hbox{E}\kern-.125emX}}

\begin{document}

\title{An Executable Benchmark for LLM-Based HLS Repair:Design Complexity and Repair Underconstraint}

\author{
\IEEEauthorblockN{Maisha Mastora}
\IEEEauthorblockA{
Department of Electrical and Computer Engineering\\
University of New Hampshire\\
Durham, NH, USA\\
maisha.mastora@unh.edu
}
\and
\IEEEauthorblockN{Dean Sullivan}
\IEEEauthorblockA{
Department of Electrical and Computer Engineering\\
University of New Hampshire\\
Durham, NH, USA\\
dean.sullivan@unh.edu
}
}

\maketitle
\begin{tikzpicture}[remember picture, overlay]
\node[anchor=north] at ([yshift=-0.4cm]current page.north)
  {\small\textit{This paper has been accepted to The 44th IEEE International Conference on Computer Design
(ICCD 2026)}};
\end{tikzpicture}

\begin{abstract}
Automated repair of High-Level Synthesis (HLS) designs using large
language models (LLMs) is an emerging but underexplored problem.
While LLM-based repair has shown strong results on register-transfer
level (RTL) Verilog, the only prior systematic study of HLS logic
repair reports just 10.5\% correction accuracy for GPT-4---with no analysis of why repair fails or what drives difficulty. In this paper, we present the first comprehensive evaluation of LLM-based HLS repair across four models (GPT-4o, GPT-4o-mini, GPT-5.4, and Claude Opus~4.6), evaluated on 125 benchmark instances spanning eight logic bug types across three open-source HLS suites (CHStone, MachSuite, and Polybench). We construct, to the best of our knowledge, the first executable
APR-style HLS repair benchmark with suite-specific functional oracles,
enabling pass@k evaluation rather than the string-match approximations
used in prior work. Our results reveal that design context and scale, rather than bug type alone, dominate repair difficulty: repair rates range from 6--45\% on complex cryptographic kernels (CHStone) to
81--93\% on compact algorithmic kernels (MachSuite) despite identical bug type distributions. We further introduce \textit{solution multiplicity}---the fraction of distinct patches generated across repeated repair attempts---as an empirical measure of repair underconstraint, and show that it strongly predicts repair failure across all four models (Spearman $\rho = -0.393$,
$p = 6.61 \times 10^{-20}$). Frontier models reduce underconstrained
 instances from 41--46 (GPT-4o, GPT-4o-mini) to just 3 (Claude
Opus~4.6), with overall pass@1 improving from 42--53\% to 73--76\%.
SHFT bugs remain consistently hard across all models, and semantic bug
classes such as buffer indexing become reliably repairable only at
frontier model scale. These findings establish a principled framework for understanding LLM repair limitations in HLS and show that future APR benchmarks must include complex, executable, and weakly identifiable designs that remain challenging after frontier LLM repair.
\end{abstract}

\begin{IEEEkeywords}
automated hardware repair, high-level synthesis, large language models,
repair identifiability, solution multiplicity, program repair benchmarks
\end{IEEEkeywords}

\section{Introduction}
\label{sec:intro}

Automated repair of hardware designs is an active research problem with practical stakes: logic bugs discovered after tape-out cost orders of magnitude more to fix than bugs caught during design. The manual hardware design debugging cycle consists of reproduce, localize, patch, and re-verify. This process consumes a substantial fraction of front-end engineering time. For designs constructed in Verilog at the register-transfer level (RTL), search-based tools such as CirFix~\cite{cirfix2022} and symbolic tools such as RTL-Repair~\cite{rtlrepair2024} demonstrate correct repair on standard benchmarks, and LLM-based approaches now achieve pass@1 rates above 50\%~\cite{arsp2025,lik2024} with continued improvement at each model generation.

High-level synthesis (HLS) raises the design abstraction level,
allowing hardware to be specified in C/C++ and compiled to RTL by a
synthesis tool. HLS is increasingly common in FPGA and ASIC workflows~\cite{xiong2024hlspilot}, and it broadens the hardware development
pathway with more flexibility. Automated debugging support
at this level raises practical importance. Yet LLM-based repair at the HLS level is barely understood. The only prior systematic study,
HLSDebugger~\cite{hlsdebugger2025}, reports that GPT-4 achieves just
10.5\% correction accuracy on HLS logic bugs under zero-shot evaluation.
No prior work explains \textit{why} repair is so difficult at this level, or what properties of a design instance drive repair success or failure.

A second motivation is benchmark adequacy. If modern LLMs already repair most instances in a benchmark, then that benchmark is no longer a useful stress test for sophisticated APR engines. This concern is especially relevant for HLS, where many public benchmarks are compact kernels with short source files and limited functional oracles. We therefore study not only whether LLMs can repair HLS bugs, but whether existing HLS
benchmark families remain difficult enough to evaluate next-generation
repair engines.

This paper addresses that gap with three contributions.

\textbf{First}, we construct, to the best of our knowledge, the first executable APR-style HLS repair benchmark with functional oracles. We build 125 validated instances spanning eight C-logic bug types across three open-source HLS suites (CHStone, MachSuite, Polybench), pairing each buggy design with a functional oracle derived from the suite's native
testing infrastructure. Prior work~\cite{hlsdebugger2025} relies on string matching against ground truth snippets; our benchmark enables pass@k evaluation, which is standard practice in software repair and avoids conflating plausible-but-incorrect patches with correct ones.

\textbf{Second}, we evaluate four LLMs---GPT-4o, GPT-4o-mini, GPT-5.4,
and Claude Opus~4.6---on all 125 instances using a uniform zero-shot
prompt and report pass@1, pass@5, and per-bug-type breakdowns. Our
results reveal that not only the bug type, but also design context and scale, significantly determine repair difficulty. Pass@1 ranges from 6--45\% on CHStone's large cryptographic kernels to 81--93\% on MachSuite's compact algorithmic kernels, despite both suites containing identical bug type
distributions drawn from the same source dataset. This finding holds
across all four models and implies that existing single-suite evaluations significantly mischaracterize LLM repair capability.

\textbf{Third}, we introduce \textit{solution multiplicity}---the
fraction of distinct patches across $N$ repair attempts---as an
empirical measure of repair underconstraint. High multiplicity on failing instances indicates the LLM is exploring without converging on a correct fix, reflecting insufficient observational constraint rather than model incapability. We show that solution multiplicity is strongly negatively correlated with repair success across all four models (Spearman $\rho = -0.393$, $p = 6.61 \times 10^{-20}$), and that frontier models nearly eliminate underconstrained instances: from 41--46 such instances for GPT-4o and GPT-4o-mini, down to 3 for Claude
Opus~4.6.

The rest of the paper is organized as follows. Section~\ref{sec:background} reviews related work. Section~\ref{sec:motivation} motivates why HLS repair is an identifiability
challenge. Section~\ref{sec:setup} describes benchmark construction and
experimental setup. Section~\ref{sec:results} presents results. Section~\ref{sec:discussion} discusses implications, and
Section~\ref{sec:conclusion} concludes.

\section{Background and Related Work}
\label{sec:background}

\subsection{Automated RTL Repair}

Automated repair for hardware designs has been focused on RTL verilog variants. Early search-based approaches such as
CirFix~\cite{cirfix2022} applied genetic programming over a mutation
operator space to repair functional bugs in Verilog designs. Subsequent
work explored symbolic methods: RTL-Repair~\cite{rtlrepair2024}
formulates repair as a constraint satisfaction problem over candidate
patch expressions, achieving substantially faster repair times on
datapath-dominated designs. SRepair~\cite{liu2025srepair} extends symbolic repair through regression-based patch ranking. ~\cite{Mastora2026DictionaryHDLRepair} presents a dictionary-guided HDL repair system that combines Another Tool for Language Recognition (ANTLR)-derived design specific mutation vocabularies with a simulation-divergence fault
localization (FL) for RTL program repair. 

More recently, LLMs have been applied directly to RTL repair.
RTLFixer~\cite{rtlfixer2024} demonstrated that LLMs can resolve RTL syntax errors with high reliability. HDLDebugger~\cite{hdldebugger2025}
extended this to functional debugging through retrieval-augmented generation. LiK~\cite{lik2024} showed that accurate bug localization
improves repair substantially, with functional pass@1 rising from
40.39\% to 58.92\% on Verilog benchmarks. ARSP~\cite{arsp2025}
introduced semantic partitioning to guide LLM repair, reporting pass@1
of 77.92\% and pass@5 of 83.88\%. CraftRTL~\cite{craftrtl2024}
demonstrated that frontier models already achieve competitive performance on Verilog repair without fine-tuning, and VeriRAG~\cite{verirag2025} showed that retrieval augmentation yields up to a 7.72-fold improvement
over zero-shot baselines.

Collectively, these results establish LLM-based RTL repair as a mature and rapidly improving area. Critically, all of these efforts operate exclusively at the RTL abstraction level. No prior work has
systematically evaluated LLM repair effectiveness on HLS logic bugs with
functional oracles. HWE-Bench~\cite{hweb2026} confirms that existing hardware repair benchmarks contain no HLS instances. This gap motivates the present work.

\subsection{LLM-Based HLS Debugging and Repair}

High-level synthesis enables hardware generation from C/C++
descriptions, broadening development to engineers without RTL
expertise~\cite{xiong2024hlspilot}. Debugging HLS designs presents
distinct challenges: the semantic gap between C++ source behavior and
synthesized hardware behavior means bugs can be non-obvious at the source level~\cite{chrysalis2024}.

Existing work on LLM-based HLS assistance has primarily targeted code generation rather than repair. HLSPilot~\cite{xiong2024hlspilot} explores feedback-driven generation with synthesis constraints, while
SynthAI~\cite{sheikholeslam2024synthai} automates pragma insertion and design refinement. For HLS repair, Xu et al.~\cite{hlsxu2023} proposed
repairing C/C++ for HLS synthesizability, targeting unsupported constructs such as recursion and dynamic memory.

The most directly relevant prior work is HLSDebugger~\cite{hlsdebugger2025},
which targets logic bug identification and correction in HLS C++ code. HLSDebugger introduces a labeled dataset of 300K HLS bug instances spanning eight logic bug types and proposes a fine-tuned encoder-decoder
architecture for simultaneous bug localization and correction. Under
zero-shot evaluation, GPT-4 achieves only 10.5\% correction accuracy on
the Chrysalis benchmark~\cite{chrysalis2024,chrysalisaspdac2024}, while their fine-tuned model reaches 37.6\%. Crucially, HLSDebugger evaluates
correction through string matching against ground truth snippets rather
than functional testbench execution, and does not investigate what structural properties drive repair difficulty. Our work is complementary: we replace direct string-match scoring with executable oracles and shift the
question from aggregate accuracy to what makes a design hard.
\section{Motivation}
\label{sec:motivation}
\subsection{Repair Identifiability and Underconstraint}

The distinction between \textit{plausible} and \textit{correct} repairs
is well established in software APR~\cite{cirfix2022}: a plausible repair passes all available tests but may not be semantically correct, a failure mode rooted in test suite incompleteness. In the hardware domain, the same mechanism applies: a weak oracle admits multiple structurally distinct patches, making the repair problem
\textit{underconstrained}. Hardware repair benchmarks typically characterize instances only by bug type, without accounting for how
strongly the oracle constrains the repair search. This work addresses
that gap for HLS by introducing \textit{solution multiplicity} as an
empirical metric of identifiability computable directly from LLM repair behavior.

\subsection{Repair Identifiability in HLS}
\label{sec:theory}

A repair instance is \textit{identifiable} if the available test observations uniquely constrain the correct patch, and \textit{underconstrained} if multiple structurally distinct patches all satisfy the oracle. While this distinction is implicit in software APR~\cite{cirfix2022}, hardware repair benchmarks have not explicitly characterized instances along this dimension. We identify three structural properties of HLS designs that make underconstraint
particularly prevalent, and that explain why single-suite evaluations
produce misleading repair rate estimates.

\textbf{Single-vector oracles.}
HLS benchmarks typically contain one fixed test input. CHStone's AES
program exercises only the 128-bit key path; bugs injected into the
192 or 256-bit key paths pass the oracle silently. This coverage gap
was the primary rejection criterion during benchmark construction: 75
candidate instances were discarded because the oracle could not
distinguish buggy from correct behavior. Single-vector coverage is
specific to HLS benchmark suites and makes underconstraint a structural property of the evaluation, not an edge case.

\textbf{Algorithmic opacity.}
HLS source is a C function body with no explicit timing, wire
assignments, or signal topology. An LLM repairing an HLS bug must infer
hardware-semantic intent from C source alone. For complex designs such as AES key scheduling or GSM channel coding, many syntactically valid edits satisfy the oracle while implementing incorrect hardware behavior, increasing the plausible-but-incorrect patch space.

\textbf{Context dilution from entry-point compilation.}
HLS tools compile from a single entry file that \texttt{\#include} all helper modules. CHStone's AES benchmark aggregates $\sim$40~KB of source into one context; the bug site is a few lines within that. Prior work identifies bug signal dilution as a key LLM failure mode in long-context repair~\cite{arsp2025}. HLS amplifies this problem because entry-point compilation is a toolchain requirement, not a design choice. Our results confirm the effect directly. CHStone instances ($\sim$40~KB) achieve 6--45\% pass@1 versus 81--93\% for MachSuite ($\sim$3~KB), despite identical bug type distributions.

These three properties compound rather than act independently. The large entry-point programs are also the cryptographic kernels hardest to infer from source and the ones whose single-vector harnesses leave the widest margin for plausible-but-wrong patches, which is why CHStone has the hard cases. We observe the combined effect empirically through
\textit{solution multiplicity} (defined in Section~\ref{sec:setup}),
which measures patch diversity across repeated repair attempts without requiring formal specification analysis or ground-truth models.

\section{Experimental Setup}
\label{sec:setup}

\begin{figure*}[!t]
\centering
\includegraphics[width=\textwidth, trim=0 100pt 0 100pt, clip]{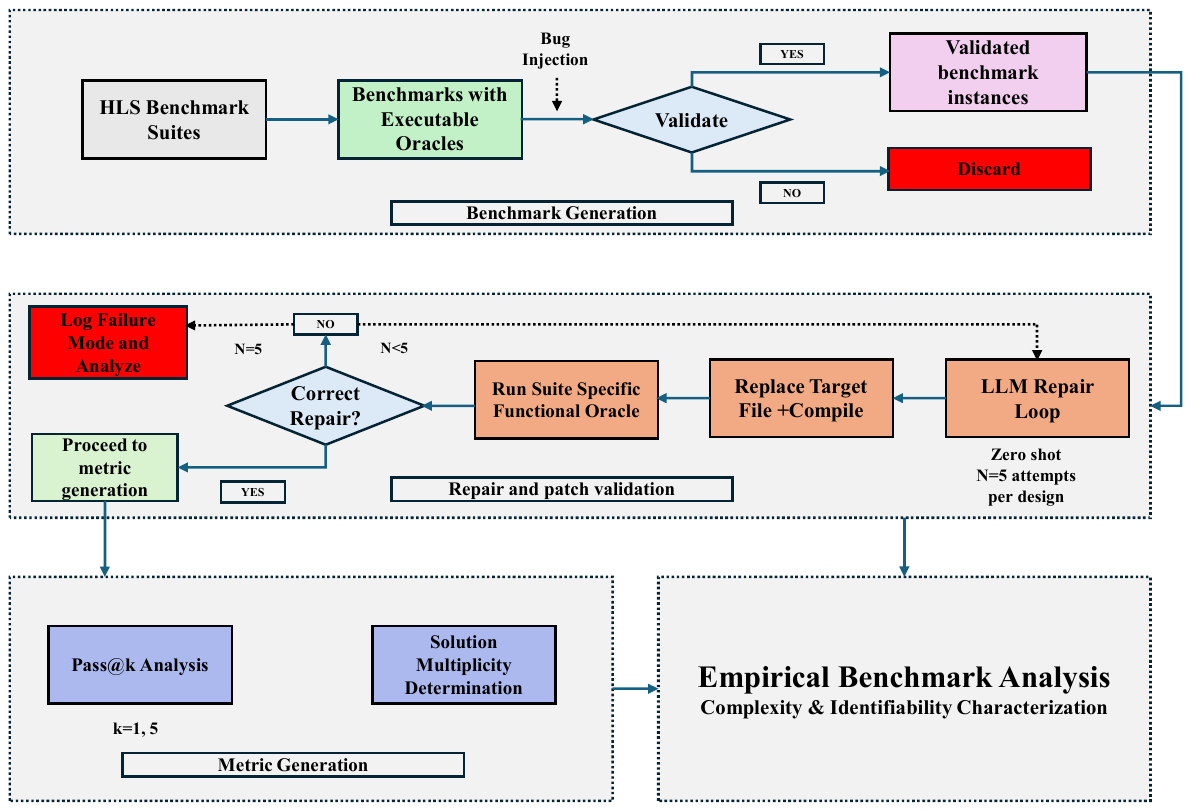}
\caption{System overview. HLS benchmark suites are paired with executable oracles and validated via bug injection to produce a benchmark of correct and buggy instance pairs. Validated instances are repaired zero-shot by LLMs (N=5 attempts per design); each candidate patch is recompiled and checked against a suite-specific functional oracle, then scored by pass@k and solution multiplicity. Failure modes and repair outcomes are aggregated for empirical complexity and identifiability characterization.}
\label{fig:system_overview}
\end{figure*}

Figure~\ref{fig:system_overview} summarizes the full workflow. The three HLS suites serve a dual role: they are the origin of the Chrysalis bug specifications and they supply the executable programs and oracles into which those bugs are injected. The injection pipeline produces 125 validated instances, which are repaired by four LLMs and evaluated through suite-specific functional oracles before metric computation and
analysis.

\subsection{HLS Benchmark Construction}

We build an executable HLS repair benchmark by pairing bug
specifications from Chrysalis-HLS~\cite{chrysalis2024} with three
open-source HLS suites: CHStone~\cite{chstone},
MachSuite~\cite{machsuite}, and Polybench~\cite{polybench}. Chrysalis provides 1,500+ function-level bug specifications across 18 bug types without oracle or testbench, only for string-match evaluation~\cite{hlsdebugger2025} and cannot verify functional correctness of a repair; we restrict to the eight C-logic-compatible types testable with
\texttt{gcc}: out-of-bounds array access (OOB), bit shift errors (SHFT), infinite loops (INF), manual loop unrolling errors (MLU), wrong zero-initialization (ZERO), unintended sign extension (USE), buffer indexing errors (BUF), and uninitialized variable reads (INIT). We filter to single-bug instances only (Chrysalis Error Size~$= 1$).

For each candidate instance, our injection pipeline: (1) copies the
benchmark source to a \texttt{correct/} directory and verifies it passes the oracle; (2) injects the Chrysalis faulty code snippet via string replace into a \texttt{buggy/} copy; (3) verifies the buggy version fails the oracle. Instances that fail verification in either direction are discarded. This two-sided check makes the benchmark executable rather than declarative: any patch that flips the oracle back to passing is a genuine functional repair under the suite's own test.

\textbf{CHStone} (41 instances). Each CHStone program has a
self-checking \texttt{main()} that sets \texttt{main\_result~$= 0$} on
correct output. Oracle: \texttt{gcc -O0 <entry>.c -lm}; pass iff exit
code~$= 0$. INF bugs are detected by a 10-second timeout. JPEG was
excluded due to duplicate-symbol compilation issues. Five programs
contributed instances: AES (17), ADPCM (16), GSM (5), DFADD (2),
Blowfish (1).

\textbf{MachSuite} (40 instances). MachSuite provides a harness-based
oracle (\texttt{check\_data()} in \texttt{local\_support.c}) that
compares output to a reference \texttt{check.data} file and returns
exit code~$-1$ on failure. Compilation: \texttt{gcc -O0 -I<common> common/support.c common/harness.c local\_support.c <kernel>.c -lm}. 10 out of 14 kernels compile cleanly; four were skipped due to compilation
issues leading to oracle failures.

\textbf{Polybench} (44 instances). Polybench has no built-in
self-check. We implement a differential oracle: compile the correct
version with \texttt{-DPOLYBENCH\_DUMP\_ARRAYS}, capture
\texttt{stderr} as \texttt{ref\_output.txt}, then compare
patched-version output byte-for-byte against reference. Reference
output is generated once per kernel and reused across all instances.
All 30 kernels compile and produce deterministic output; 27 have
matching Chrysalis bug specifications.

The final benchmark contains 125 instances across eight bug types
(Table~\ref{tab:bugdist}). BUF and INIT are underrepresented because
their bug sites tend to fall in code paths not covered by single test
vectors.

\begin{table}[!t]
\centering
\caption{Bug type distribution across 125 benchmark instances.}
\label{tab:bugdist}
\begin{tabular}{lrrrr}
\toprule
Bug Type & CHStone & MachSuite & Polybench & Total \\
\midrule
MLU  & 7  & 10 & 9  & 26 \\
ZERO & 7  & 7  & 10 & 24 \\
INF  & 7  & 5  & 10 & 22 \\
OOB  & 7  & 8  & 3  & 18 \\
USE  & 4  & 3  & 10 & 17 \\
SHFT & 7  & 1  & 0  & 8  \\
INIT & 1  & 5  & 1  & 7  \\
BUF  & 1  & 1  & 1  & 3  \\
\midrule
Total & 41 & 40 & 44 & 125 \\
\bottomrule
\end{tabular}
\end{table}

\subsection{LLM Repair Pipeline}

All models receive a uniform zero-shot prompt: \textit{``The following C
source file contains bugs. Fix all of them and return the complete
corrected file.''} No hints about bug location or type are provided.
The full buggy source file is passed as the user message; the model
returns a complete replacement file. We run $N = 5$ independent repair
attempts per instance at temperature 1.0. Token budgets use a per-instance dynamic allocation
(min 1024, max 16384 tokens) based on source file length. We report
model names as used in our experiment logs; exact API endpoint
identifiers and evaluation dates are included in the artifact metadata, which will be released for public availability upon the acceptance of the paper. 

We evaluate four models: \textbf{GPT-4o} (OpenAI, standard generation),
\textbf{GPT-4o-mini} (OpenAI, cost-reduced), \textbf{GPT-5.4} (OpenAI
Batch API), and \textbf{Claude Opus~4.6} (Anthropic Batch API).
GPT-4o and GPT-4o-mini represent the generation-4 frontier; GPT-5.4
and Opus~4.6 represent the current generation-5 frontier. The prompt is deliberately minimal to observe repair identifiability---whether the oracle and source together constrain the patch without any external information, so a fixed zero-shot prompt isolates instance and model-level variation. This establishes the baseline on which guided methods can later be evaluated.

\subsection{Metrics}

\textbf{pass@k} uses the standard unbiased estimator from Chen et
al.~\cite{chen2021evaluating}:
\begin{equation}
\text{pass@}k = 1 - \frac{\binom{n - c}{k}}{\binom{n}{k}}
\end{equation}
where $n = 5$ is the total runs per instance and $c$ is the number of
passing runs. pass@1 is the primary metric; pass@5 gives an upper bound
on what is achievable with a fixed sample budget.

\textbf{Solution multiplicity} is defined as the number of distinct
patches divided by the total number of runs ($n = 5$). Two patches are
distinct if they differ in any character after stripping whitespace.
Multiplicity $= 1.0$ means every run produced a different patch;
multiplicity $< 0.5$ indicates convergence on a small set of repairs.
Because multiplicity is model- and decoding-dependent, we interpret it
as an empirical repair-behavior metric rather than an intrinsic property
of the benchmark instance alone.

\textbf{Failure mode} categorizes each non-passing patch as: compile
failure (patch does not compile), test failure (compiles but oracle
fails), or timeout (INF-type bugs that do not terminate).

\section{Results}
\label{sec:results}

\subsection{Overall Repair Performance}

Table~\ref{tab:overall} reports pass@1 and pass@5 for all four models
across all 125 instances. There is a clear generational progression:
GPT-4o-mini (42.4\% pass@1) $<$ GPT-4o (53.3\%) $<$ GPT-5.4 (73.4\%)
$\approx$ Opus~4.6 (75.7\%). This 33-percentage-point spread confirms
that model capability significantly influences HLS repair success,
consistent with findings in RTL repair~\cite{arsp2025,lik2024}.

\begin{table}[!t]
\centering
\caption{Overall pass@1, pass@5, and solution multiplicity by model.}
\label{tab:overall}
\begin{tabular}{lrrr}
\toprule
Model & pass@1 & pass@5 & Multiplicity \\
\midrule
GPT-4o-mini     & 42.4\% & 63.2\% & 0.970 \\
GPT-4o          & 53.3\% & 70.4\% & 0.954 \\
GPT-5.4         & 73.4\% & 78.4\% & 0.771 \\
Claude Opus~4.6 & 75.7\% & 78.4\% & 0.310 \\
\bottomrule
\end{tabular}
\end{table}

The reduction in multiplicity is as important as the increase in pass
rate. Weaker models generate highly diverse patches across attempts,
while Opus~4.6 often converges to the same patch. This indicates that
stronger models are not only more accurate but also more stable in the
repair space.

\subsection{Design Complexity Dominates Repair Difficulty}

\textbf{RQ1: Does design complexity predict repair difficulty more
strongly than bug type?}

Table~\ref{tab:suite} breaks pass@1 down by suite. The suite-level
variation is dramatic and consistent across all models: CHStone achieves
13.7--44.9\%, while MachSuite achieves 48.5--92.5\% and Polybench
56.8--90.9\%. Critically, all three suites draw bugs from the same
Chrysalis specifications and contain the same eight bug types---the only
structural difference is design complexity and file size ($\sim$40~KB
for CHStone vs.\ $\sim$3--4~KB for MachSuite/Polybench).

\begin{table}[!t]
\centering
\caption{pass@1 by suite and model. The CHStone vs.\ MachSuite gap holds
across all models despite identical bug type distributions.}
\label{tab:suite}
\begin{tabular}{lrrrr}
\toprule
Suite & GPT-4o-mini & GPT-4o & GPT-5.4 & Opus~4.6 \\
\midrule
CHStone   & 21.0\% & 13.7\% & 37.6\% & 44.9\% \\
MachSuite & 48.5\% & 74.0\% & 92.5\% & 90.5\% \\
Polybench & 56.8\% & 71.4\% & 89.5\% & 90.9\% \\
\bottomrule
\end{tabular}
\end{table}

The MachSuite--CHStone gap ranges from 27.5 percentage points
(Opus~4.6) to 60.5 points (GPT-4o), answering RQ1 affirmatively:
suite complexity dominates bug type as a repair difficulty predictor.
Because the bug-type mix is held fixed across suites by construction,
this gap cannot be attributed to one suite containing intrinsically
harder bug categories. MachSuite and Polybench are essentially solved by
frontier models ($>$90\%); CHStone remains hard for all models.

\begin{figure}[!t]
\centering
\safeincludegraphics[width=\linewidth]{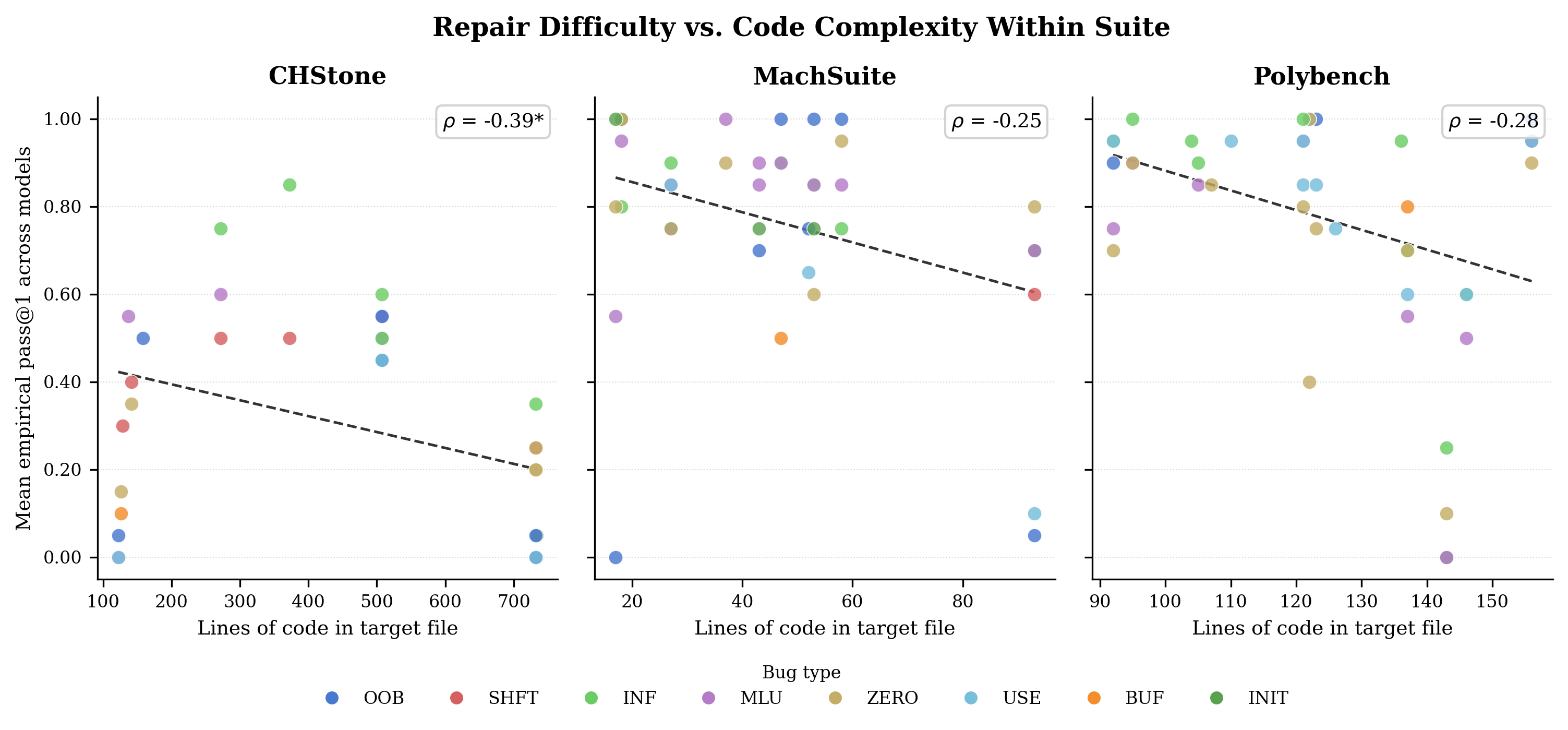}
\caption{Repair difficulty versus target-file LOC within each suite.
Each point is one benchmark instance, colored by bug type. The y-axis
reports mean empirical pass@1 across models. CHStone shows a significant
negative correlation ($\rho=-0.389$, $p=0.012$); MachSuite and Polybench
show weaker non-significant negative trends due to narrower LOC ranges.}
\label{fig:complexity_scatter}
\end{figure}

Figure~\ref{fig:complexity_scatter} further shows that complexity is
not only a between-suite effect. Within CHStone, target-file LOC is
significantly negatively correlated with empirical pass@1 ($\rho =
-0.389$, $p = 0.012$), confirming that repair difficulty scales
continuously with file size even within a single suite. In MachSuite
and Polybench, where file sizes are more uniform, within-suite
correlations do not reach significance, consistent with range
restriction rather than absence of effect.

\subsection{Solution Multiplicity and Underconstraint}

\textbf{RQ2: Does solution multiplicity predict repair failure?}

Solution multiplicity is strongly negatively correlated with pass@1
across all 125 instances (Spearman $\rho = -0.393$,
$p = 6.61 \times 10^{-20}$). Instances where the model produces a
different patch on every run are overwhelmingly failures; instances
where the model converges on one or two patches are overwhelmingly
passes. The correlation is computed over the full dataset, so it does
not depend on any single suite or bug category.

Model-level multiplicity tracks capability inversely: GPT-4o-mini
(0.970) $>$ GPT-4o (0.954) $>$ GPT-5.4 (0.771) $>$ Opus~4.6 (0.310).
Opus~4.6's multiplicity of 0.310 is the lowest of any model: it
produces the same correct patch on all 5 runs for instances it can
repair, and similarly converges (incorrectly) on failing instances,
reflecting high model confidence.

We define an \textit{underconstrained} instance as one where
multiplicity $\geq 0.8$ and pass@5~$= 0$: the model generates maximally
diverse patches and none pass. Table~\ref{tab:undercon} shows the count
by model and suite. Older models have 35--46 such instances,
concentrated on CHStone; frontier models reduce this to 3 (Opus~4.6).
The 3 remaining underconstrained instances for Opus~4.6 involve large
cryptographic files or complex loop structures where the oracle covers
only a subset of code paths.

\begin{table}[!t]
\centering
\caption{Underconstrained instances (multiplicity $\geq 0.8$,
pass@5~$= 0$) by model and suite.}
\label{tab:undercon}
\begin{tabular}{lrrrr}
\toprule
Model & CHStone & MachSuite & Polybench & Total \\
\midrule
GPT-4o-mini     & 23 & 12 & 10 & 46 \\
GPT-4o          & 25 &  5 &  5 & 35 \\
GPT-5.4         & 20 &  0 &  3 & 23 \\
Opus~4.6        &  2 &  1 &  0 &  3 \\
\bottomrule
\end{tabular}
\end{table}

Multiplicity also separates two failure modes that a pass-rate-only view
would merge. On the 3 BUF instances, GPT-4o records multiplicity 0.800
with 0.0\% pass@1: it produces similar wrong patches every run, a
confident-but-incorrect convergence. SHFT shows the opposite signature
(multiplicity 1.000, 5.0\% pass@1 for GPT-4o): a different guess each
run, the hallmark of pure underconstraint. The same zero pass rate thus
arises from two distinct causes, and multiplicity is what tells them
apart.

\subsection{Bug-Category Analysis}

\textbf{RQ3: Which bug types are consistently hard, and which improve
with model scale?}

\begin{figure}[!t]
\centering
\safeincludegraphics[width=\linewidth]{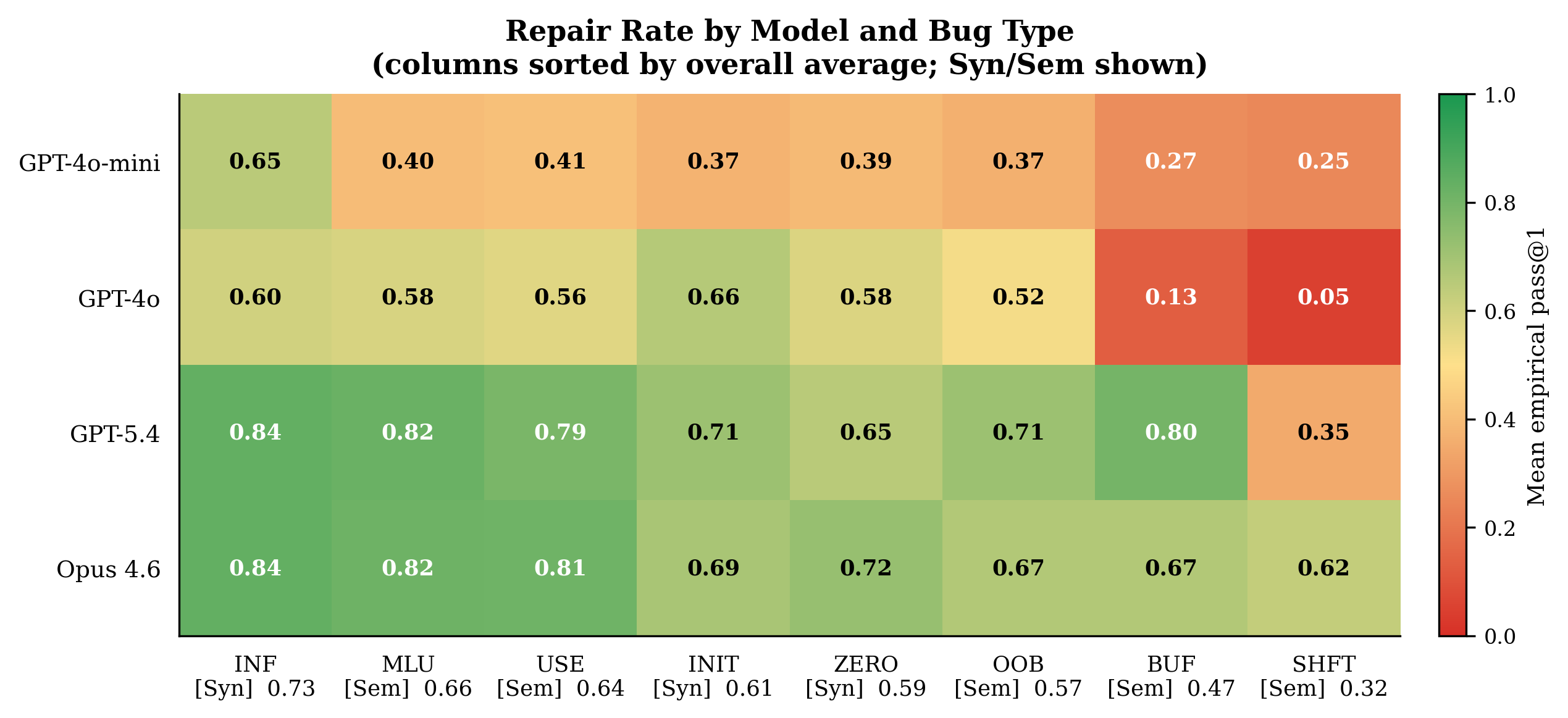}
\caption{Repair rate by model and bug type. Columns sorted by overall
average empirical pass@1. Model scaling improves most bug types, but
SHFT remains the hardest category. BUF shows strong improvement in
larger models, suggesting some semantic bug classes become repairable
only at frontier scale.}
\label{fig:model_bug_heatmap}
\end{figure}

Figure~\ref{fig:model_bug_heatmap} shows empirical pass@1 by model and
bug type. Model scaling improves most bug types, but SHFT remains the
hardest category and buffer-indexing bugs improve only with frontier
models.

Four patterns are notable. (1)~\textbf{INF is consistently easiest}:
loop termination conditions are syntactically localized and semantically
obvious across all suites and models. (2)~\textbf{SHFT is consistently
hardest}: correct shift amounts require bit-level precision the model
rarely recovers without localization hints; even Opus~4.6 achieves only
62.5\%, and SHFT carries enough instances ($n=8$) for this to be a
stable trend. (3)~Among the smaller categories, \textbf{BUF ($n=3$)
illustrates a semantic capability threshold}: GPT-4o repairs none of its
instances while the generation-5 models repair most, consistent with
buffer-indexing semantics being learnable but frontier-dependent; we
read this as illustrative rather than a precise rate given the category
size. (4)~\textbf{MLU is suite-dependent}: GPT-4o achieves 2.9\% on
CHStone MLU but 55.4\% overall, because MachSuite and Polybench MLU
instances are in small files where the loop structure is clear. The
aggregate and suite-level conclusions rest on the full dataset and do
not depend on the smallest per-type cells.

\begin{figure}[!t]
\centering
\safeincludegraphics[width=\linewidth]{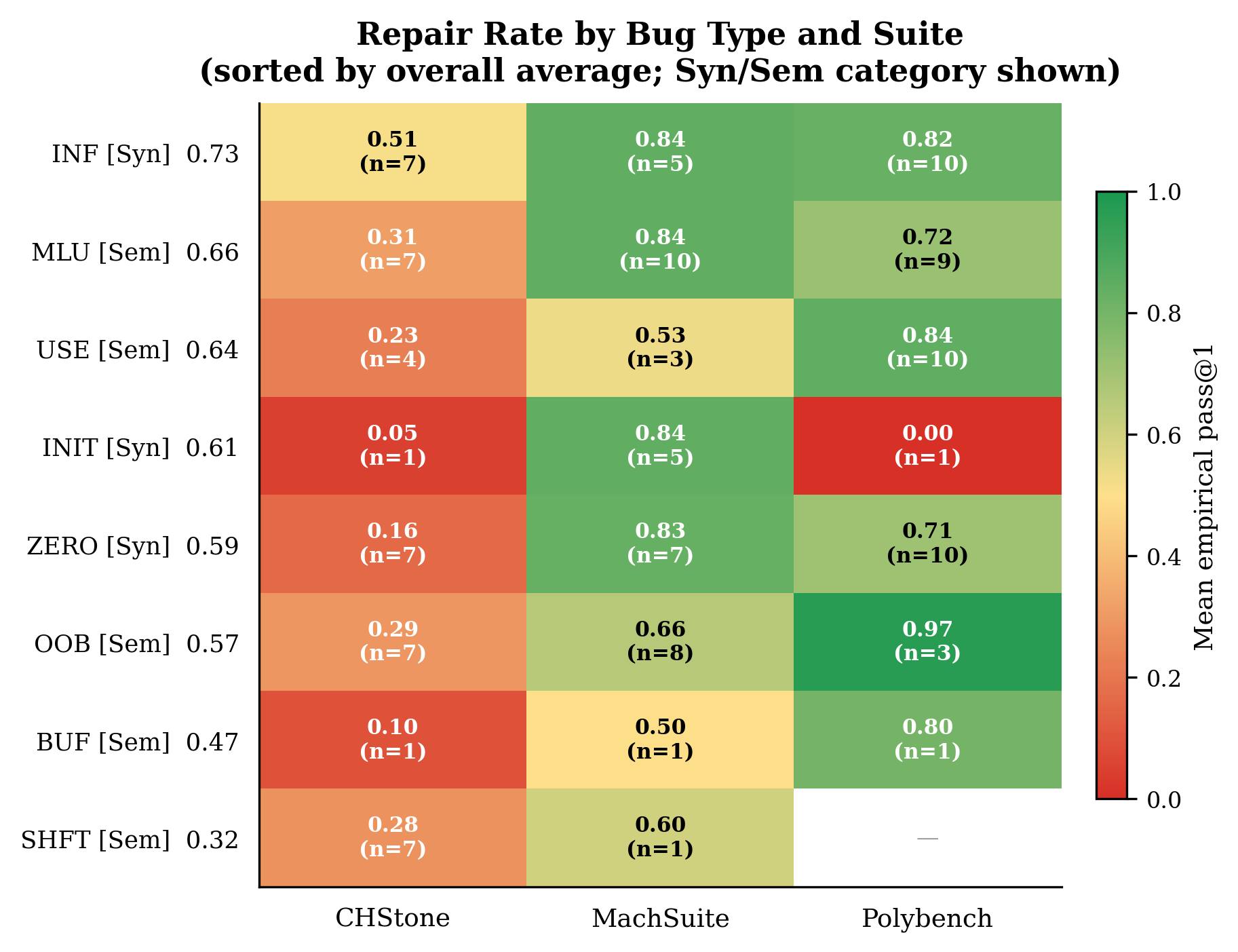}
\caption{Repair rate by bug type and suite. Each cell reports mean
empirical pass@1 across models, with the number of instances in
parentheses. The same bug type can have very different repair rates
across suites, confirming that benchmark context substantially mediates
bug-type difficulty.}
\label{fig:bug_suite_heatmap}
\end{figure}

Figure~\ref{fig:bug_suite_heatmap} shows empirical pass@1 by bug type
and suite, illustrating how the same bug type can carry very different
repair rates depending on design context.

\textbf{pass@k gain by bug type.} SHFT benefits most from additional
attempts ($+$15--37\% from pass@1 to pass@5), confirming the correct
shift value can occasionally be sampled but is not reliably produced.
USE plateaus early ($+$3.5\% for GPT-4o), indicating the model either
knows the sign-extension fix or does not. BUF at 0.0\% for GPT-4o
shows zero gain at any~$k$---not a sampling problem but a semantic
capability gap that only larger models overcome.

\subsection{Structural Metrics and Solution Multiplicity}

To compare alternative predictors of repair difficulty, we analyze
target-file LOC, cyclomatic complexity (CCN)~\cite{mccabe1976complexity},
and solution multiplicity. Following established practice for
nonparametric evaluation in SE research~\cite{arcuri2011practical}, we
use Spearman rank correlation ($\rho$) throughout.

LOC is the strongest predictor ($\rho = -0.564$, $p < 0.001$),
primarily reflecting between-suite scale differences. Cyclomatic
complexity yields a non-significant positive correlation ($\rho =
+0.174$, $p = 0.052$), the \textit{opposite} of the structural
complexity hypothesis. CCN is near-uniform across suites (avg.\ 6.95,
6.92, and 6.36 for CHStone, MachSuite, and Polybench respectively),
confirming this is not a suite-level confound: CHStone files are large
but not structurally more branchy. This is consistent with recent
findings that classical metrics show no consistent correlation with LLM
difficulty after controlling for length~\cite{xie2026rethinking}.
Solution multiplicity achieves $\rho = -0.393$ ($p < 0.001$), capturing
repair difficulty that CCN cannot. LOC and multiplicity thus capture
complementary dimensions: structural scale and semantic underconstraint.

\begin{figure}
\centering
\safeincludegraphics[width=\linewidth]{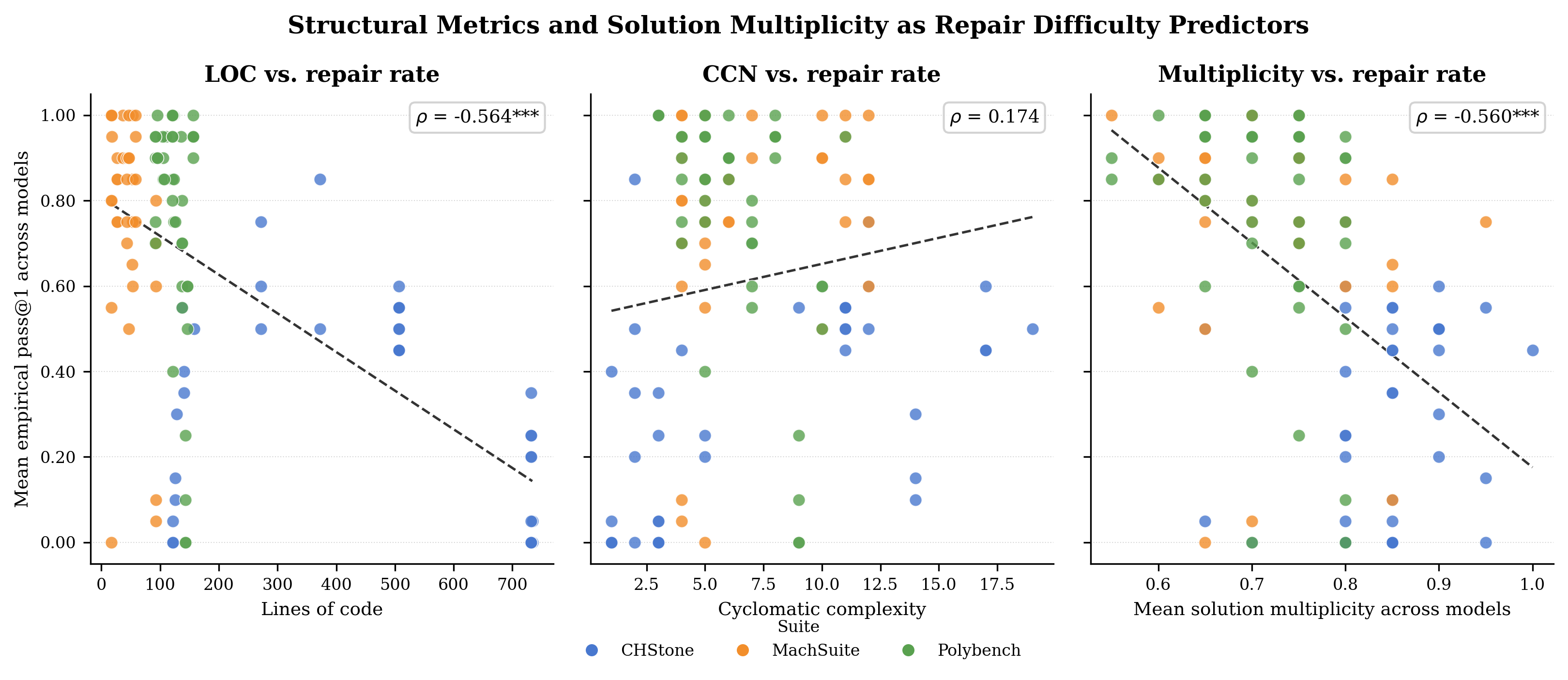}
\caption{Comparison of repair difficulty predictors. LOC and mean
solution multiplicity are both strongly negatively correlated with mean
empirical pass@1, while cyclomatic complexity is not a significant
negative predictor.}
\label{fig:predictor_comparison}
\end{figure}

\textbf{Syntactic vs.\ semantic bug categories.}
We classify INF, INIT, and ZERO as syntactic and SHFT, BUF, OOB, MLU,
and USE as semantic bugs. For weaker models a visible gap exists
(GPT-4o-mini: 0.47 vs.\ 0.34; GPT-4o: 0.61 vs.\ 0.37), but it narrows
sharply for frontier models (GPT-5.4: 0.73 vs.\ 0.69; Opus~4.6: 0.75
vs.\ 0.72). A Mann-Whitney U test does not reach significance
($p = 0.159$). The syntactic/semantic distinction describes a capability
threshold that frontier models have largely
overcome~\cite{chen2025bugfix,ouyang2024benchmarking} rather than a
complexity-independent predictor. 
Figure~\ref{fig:predictor_comparison}
summarizes the three predictor correlations.

\section{Analysis and Discussion}
\label{sec:discussion}

\textbf{Benchmark saturation and the need for harder APR tasks.}
Our results make a strong methodological argument: reporting repair
rates on a single HLS suite produces misleading estimates of LLM
capability and may overstate progress in APR. The CHStone-only repair
rate for GPT-4o (13.7--25.9\%) and the MachSuite rate (74--92.5\%)
would support opposite conclusions about HLS repair difficulty. More
importantly, MachSuite and Polybench are largely saturated by frontier
models, with repair rates above 90\% in several settings. If an LLM can
already solve most instances in such suites under a zero-shot prompt,
then these suites alone cannot serve as strong benchmarks for more
sophisticated APR engines. The true picture is that difficulty is driven
by design complexity---file size, algorithmic depth, and oracle
coverage---which varies more within HLS benchmarks than across bug
types. A useful HLS APR benchmark should therefore not be defined by
bug-type coverage alone. It should include executable functional oracles
rather than string matching, span both compact kernels and larger
entry-point programs, retain only oracle-observable bugs, include
instances that remain hard for frontier LLMs, and report metadata such
as LOC, oracle structure, failure mode, and identifiability metrics such
as solution multiplicity.

\textbf{Construction discipline and evaluation criterion.}
The 125-instance suite was built under a strict open-source,
no-proprietary-tool constraint. Candidate Chrysalis-HLS specifications
were retained only when the original program passed its oracle, the
injected program compiled, and the injected bug was observable; the rest
were rejected. The suite is therefore a systematically filtered
executable subset whose every member carries a guaranteed
pass/fail-discriminating test, which is what licenses the pass@k
semantics used throughout. We score repair at the C level using each
suite's executable oracle---the standard correctness criterion in
software APR---so the object of study is C-level functional repair:
producing a source patch that restores externally observable behavior.
This is the level at which HLS bugs are written, at which Chrysalis
specifications are defined, and at which a repair must first be correct,
making it the natural foundation for later work that layers
post-synthesis equivalence checks onto the same harness.

\textbf{Solution multiplicity as an LLM-native metric.}
Standard metrics (pass@k, compile rate, test fail rate) measure
outcomes. Solution multiplicity measures the repair process: is the
model converging on an answer, or exploring randomly? Our correlation
results ($\rho = -0.393$) show that multiplicity is a strong early
signal of repair difficulty that is practically useful without
ground-truth labels. Given a new benchmark instance, running 5 repair
attempts and computing multiplicity provides an immediate estimate of
whether the instance is identifiable by the model. High multiplicity
($\geq 0.8$) with no passing runs is a reliable flag for oracle
coverage gaps or context length problems.

\textbf{Failure mode analysis.}
Compile failures are stable at 11--20\% regardless of suite or bug
type, reflecting consistent LLM hallucination of invalid C syntax. Test
failures (plausible-but-incorrect patches) are complexity-dependent:
54.1\% on CHStone vs.\ much lower on simple kernels. This aligns with
software APR findings that test suite coverage drives
plausible-repair prevalence: CHStone's single hardcoded test vector
accepts a wide range of wrong patches.

\textbf{Frontier model saturation.}
GPT-5.4 and Opus~4.6 are tied at pass@5 (78.4\%) and both saturate
MachSuite/Polybench above 90\%. The marginal gain from pass@1 to
pass@5 shrinks for frontier models, indicating that they either know the
fix early or consistently fail. CHStone remains unsaturated at 51.2\%
pass@5 for Opus~4.6, making it the relevant hard benchmark for future
work.

\textbf{Memorization.}
CHStone, MachSuite, and Polybench are public benchmarks and may appear in
LLM training data. Memorization of original source would, however, raise
repair rates roughly uniformly, whereas our results show strong variation
by suite and bug type; for example, SHFT remains hard even in simpler
suites. This pattern is consistent with repair reasoning that scales with
design context rather than retrieval of memorized fixes.

\textbf{Reproducibility and implications for repair tools.}
The benchmark is organized so that future methods can be evaluated
without reimplementing the validation logic: each instance provides a
correct version, a buggy version, the target file, and an executable
oracle, and a repair system need only emit a replacement file for the
script to rebuild, run the oracle, and record the outcome. This
separation lets LLM-only, symbolic, localization-guided, and hybrid APR
methods be compared under one executable criterion. It also points to
where effort is best spent: compact kernels are near-saturated, while
CHStone-like designs will need bug localization to reduce context
dilution, execution feedback to reject plausible-but-wrong patches, and
stronger oracles to expose defects hidden by single-vector tests. The
benchmark thus identifies not only which instances are hard but which
capabilities future APR systems must add beyond zero-shot generation.

\textbf{Threats to Validity.}
\textit{Internal validity}: the primary threat is oracle sensitivity.
Bugs in code paths not exercised by single-vector tests are invisible
to the oracle; the 75 rejected candidate instances were predominantly
oracle-invisible for this reason. 
\textit{External validity}: CHStone, MachSuite, and Polybench are
widely cited public benchmarks that may appear in LLM training data,
introducing a potential memorization confound. However, memorization
would raise repair rates uniformly, whereas our results show strong
structured variation by suite and bug type. We evaluate C-level
functional repair only; whether patches remain correct after HLS
synthesis is a downstream question beyond the scope of this benchmark.
\textit{Construct validity}: we evaluate zero-shot repair exclusively.
Localization-guided or iterative repair could substantially improve
results on hard instances, particularly CHStone; the benchmark harness
is designed to support such evaluations directly. 

\textbf{Future Work.} Three directions follow directly from this work. First, the benchmark harness supports drop-in evaluation of localization-guided and iterative repair methods; applying ARSP~\cite{arsp2025} or LiK~\cite{lik2024}-style fault localization to CHStone designs is a natural next step. Second,
oracle strengthening---expanding single-vector CHStone tests to cover
additional key-length paths---would reduce underconstraint and produce
a harder, more discriminating benchmark. Third, the current suite was
constructed under a strict open-source, \texttt{gcc}-only constraint;
incorporating proprietary HLS toolchains such as Xilinx Vitis HLS or
Intel oneAPI HLS would enable synthesis-aware oracles that validate
not only C-level functional correctness but also synthesizability,
timing closure, and resource utilization, producing richer and more
realistic benchmarks for next-generation APR evaluation.
\section{Conclusion}
\label{sec:conclusion}

We presented the first comprehensive evaluation of LLM-based HLS bug
repair, built on 125 executable benchmark instances with functional
oracles spanning three open-source HLS suites and eight logic bug
types. Our main findings are: (1) design context and scale, rather than
bug type alone, dominate repair difficulty; (2) solution multiplicity
strongly predicts repair failure and provides an LLM-native measure of
oracle underconstraint; and (3) frontier models (GPT-5.4, Claude
Opus~4.6) largely solve simple HLS kernels but leave complex
cryptographic designs substantially unrepaired. SHFT bugs remain the
hardest across all models, while semantic bug classes such as buffer
indexing become reliably repairable only at frontier scale.

These results have direct implications for benchmark design: a
meaningful HLS repair benchmark must sample across design complexity,
not only bug type. They also motivate oracle-guided repair strategies
(fault localization, iterative feedback) as the next step for improving
repair on hard instances where zero-shot LLMs consistently fail to
converge.

\bibliographystyle{IEEEtran}
\bibliography{references}

\end{document}